\documentclass[namedreferences]{solarphysics}

\usepackage[hyperref,optionalrh,natbib]{spr-sola-addons} 
\usepackage{graphicx}        
\usepackage{amssymb}        
\usepackage{color}           
\usepackage{breakurl}        

\usepackage{enumerate}

\newcommand{\pder}[2]{\frac{\partial #1}{\partial #2} }

\newcommand{\aap}{    {\it Astron. Astrophys.}}

\newcommand{\apj}{    {\it Astrophys. J.}}
\newcommand{\apjl}{   {\it Astrophys. J. Lett.}}

\newcommand{\mnras}{  {\it Mon. Not. Roy. Astron. Soc.}}

\newcommand{\solphys}{{\it Solar Phys.}}
 
\newcommand{\ssr}{    {\it Space Sci. Rev.}}

\begin{document}

\begin{article}

\begin{opening}

\title{Source of a Prominent Poleward Surge During Solar Cycle 24}

\author{A.R.~\surname{Yeates}$^{1}$\sep
        D.~\surname{Baker}$^{2}$\sep
        L.~\surname{van~Driel-Gesztelyi}$^{2, 3, 4}$      
       }
\runningauthor{A.R. Yeates \textit{et al.}}
\runningtitle{Source of a Prominent Poleward Surge}

   \institute{$^{1}$ Department of Mathematical Sciences, Durham University, Durham, DH1 3LE, UK
                     email: \url{anthony.yeates@durham.ac.uk}\\ 
              $^{2}$ UCL-Mullard Space Science Laboratory, Holmbury St Mary, Dorking, Surrey, RH5 6NT, UK \\
              $^{3}$ Observatoire de Paris, LESIA, UMR 8109 (CNRS), 92195, Meudon Principal Cedex, France \\
               $^{4}$ Konkoly Observatory, Research Centre for Astronomy and Earth Sciences, Hungarian Academy of Sciences, Budapest, Hungary               
             }

\begin{abstract}
As an observational case study, we consider the origin of a prominent poleward surge of leading polarity, visible in the magnetic butterfly diagram during Solar Cycle 24. A new technique is developed for assimilating individual regions of strong magnetic flux into a surface flux transport model. By isolating the contribution of each of these regions, the model shows the surge to originate primarily in a single high-latitude activity group consisting of a bipolar active region present in Carrington Rotations 2104\,--\,05 (November 2010\,--\,January 2011) and a multipolar active region in Rotations 2107\,--\,08 (February\,--\,April 2011). This group had a strong axial dipole moment opposed to Joy's law. On the other hand, the modelling suggests that the transient influence of this group on the butterfly diagram will not be matched by a large long-term contribution to the polar field, because of its location at high latitude. This is in accordance with previous flux transport models.
\end{abstract}
\end{opening}

\section{Introduction}

The gradual transport of magnetic flux from decaying active regions on the Sun's surface toward the polar regions is essential for explaining the observed pattern of surface magnetic field \citep{1961ApJ...133..572B,1964ApJ...140.1547L,1989Sci...245..712W, 2012LRSP....9....6M, 2014SSRv..tmp...28P, 2014SSRv..tmp...43J}. However, the net magnetic flux transported to the poles plays an even more significant role in the popular Babcock--Leighton solar cycle model \citep{1961ApJ...133..572B, 1969ApJ...156....1L}. In this picture, the average latitudinal offset between positive and negative active region polarities \citep[known as Joy's law:][]{1919ApJ....49..153H} leads to an imbalance between the overall poleward transport of positive and negative polarities in each hemisphere. This forms a net poloidal magnetic field, and differential rotation of this field in the underlying convection zone is believed to produce the azimuthal magnetic field driving the subsequent solar cycle of sunspot emergence. This viewpoint is supported by observed correlations between the polar field at solar minimum and the strength of the subsequent solar cycle \citep[\textit{e.g.}][]{2013ApJ...767L..25M}. Unfortunately, our ability to predict future solar cycles remains limited by our ability to predict differences in the poloidal field production from one cycle to the next. This motivates continuing study of surface-flux transport on the Sun.

\begin{figure}[h]
\centering
\includegraphics[width=0.8\textwidth]{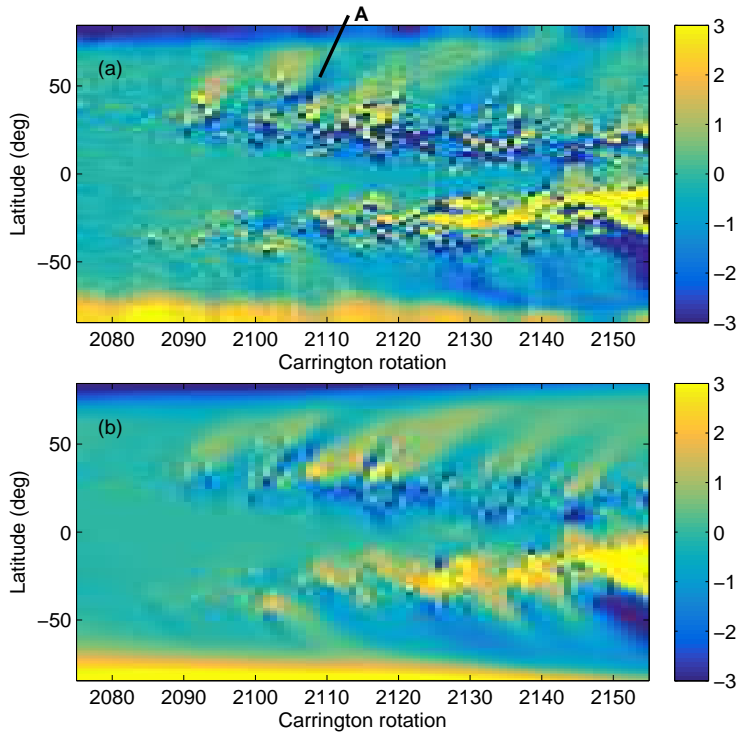}
\caption{Longitude-averaged $B_r$ as a function of time and latitude, in (a) the sequence of observed GONG synoptic maps and (b) the flux-transport simulation (Section \ref{sec:model}) for the period CR 2074\,--\,CR 2154 (August 2008\,--\,August 2014). Both figures are saturated at $\pm 3\,{\rm G}$. The leading-polarity (negative) surge A is indicated in panel (a).}
\label{fig:bfly}
\end{figure}

It has been recognised since the work of \cite{1981SoPh...74..131H} that the transport of magnetic flux from decaying active regions to the poles takes place not smoothly, but through episodic ``surges''. This is seen by inspecting the mid-latitudes in Figure \ref{fig:bfly}a, which shows the so-called butterfly diagram of longitude-averaged radial magnetic field on the photosphere, for Solar Cycle 24. The data used here are radial-component [$B_r$] synoptic maps from the \textit{Global Oscillation Network Group} (GONG, \url{http://gong.nso.edu/data/magmap/}). In particular, we focus in this article on a particular poleward surge of negative polarity in the northern hemisphere, labelled A in Figure \ref{fig:bfly}a. This surge had the leading polarity for this hemisphere, meaning the polarity of (the majority of) equatorward spots in bipolar groups.

In Section \ref{sec:model}, we describe the data-driven surface-flux transport model used to analyse the origin of surge A. The longitude-averaged $B_r$ for the simulation is shown in Figure \ref{fig:bfly}b. In fact, polar surges were important in the historical development of the surface-flux transport model \citep{2005LRSP....2....5S}, because they represent strong evidence for the presence of a poleward meridional flow.  This was suggested by \cite{1981SoPh...74..131H}, and demonstrated conclusively by \cite{1989ApJ...347..529W}, who used surface-flux transport simulations to model Solar Cycle 21. Their model reproduced surges only when they included a poleward meridional flow in addition to supergranular diffusion.

The timing of the poleward surges is dependent both on the flux-transport parameters and, of course, on fluctuations in the underlying pattern of flux emergence. For example, using both Kitt Peak and Mount Wilson Observatory magnetograms from 1974 to 2012, \cite{2012SoPh..281..577P} showed that the yearly averaged magnetic-field strength around $50^\circ$ latitude closely tracks a measure of total active region flux, provided that the latter is weighted by the latitude-difference between centroids of positive and negative flux. \cite{2012IAUS..286...88Z} developed an earlier idea of \cite{1938IzPul..16...36G} that the 11-year solar cycle consists of two or more superimposed activity impulses, different in each hemisphere. They suggested that these impulses produce poleward surges of trailing polarity in each hemisphere, while allowing waves of leading polarity (such as our surge A) to occur in the gaps between impulses. The impulses themselves may correspond to the complexes of activity studied by \cite{1983ApJ...265.1056G} and \cite{2008ApJ...686.1432G}, which comprise multiple emerging active regions at the same location over three to six solar rotations. Observations suggest that these impulses are also evident in coronal extreme-ultraviolet and soft X-ray emission \citep{2003SoPh..216..325B}. In Section \ref{sec:source}, we show that the leading-polarity surge A originated primarily from a single active region during Carrington Rotation 2107. It is better interpreted as falling between two successive impulses, rather than itself being associated with a long-lived complex of activity.

A common property of poleward surges is that they tend to occur in pairs of opposite polarity. This is clearly seen in Figure 8 of \cite{2013ApJ...768..189U}, who plot $\mathrm{d}B_r/\mathrm{d}t$ from Mount Wilson Observatory measurements and see a pattern that they call ``ripples''. The successive arrival of alternating polarity surges causes significant short-term oscillations in the polar field. When the contribution of an individual active region is isolated (as we will illustrate in Section \ref{sec:polar}), it tends to contribute to a poleward surge of each polarity, typically offset in time. Later in the cycle, these contributions largely cancel one another \citep{1989ApJ...347..529W}. As we will discuss further in Section \ref{sec:polar}, the long-term impact on the polar field then depends on the imbalance between these contributions, which is ultimately equivalent to the imbalance of net magnetic flux from the region in each hemisphere, which in turn results from cross-hemispheric flux transport and magnetic cancellation \citep{1985AuJPh..38.1045G,2004SoPh..222..345D,2013A&A...557A.141C}.

\section{Data-driven Flux Transport Model} \label{sec:model}

We model the global photospheric radial magnetic field, [$B_r(\theta,\phi,t)$], during Solar Cycle 24 using a surface-flux transport model \citep{2005LRSP....2....5S,2012LRSP....9....6M,2014SSRv..tmp...43J}. In particular, we write 
\begin{equation}
B_r = \frac{1}{{R}_\odot\sin\theta}\left(\frac{\partial}{\partial\theta}\Big(\sin\theta A_\phi\Big) - \frac{\partial A_\theta}{\partial\phi} \right),
\end{equation}
in spherical polar coordinates $(r,\theta,\phi)$, and evolve the vector potential [$(A_\theta,A_\phi)$] numerically on $r={R}_\odot$ with a finite-difference method through the equations
\begin{eqnarray}
\pder{A_{\theta}}{t} &=&  \omega(\theta){R}_\odot\sin\theta B_{r} - \frac{D}{{R}_\odot\sin\theta}\pder{B_{r}}{\phi} + S_\theta(\theta,\phi,t), \label{eqn:ath}\\
\pder{A_{\phi}}{t} &=& -u_\theta B_{r} + \frac{D}{{R}_\odot}\pder{B_{r}}{\theta} + S_\phi(\theta,\phi,t). \label{eqn:aph}
\end{eqnarray}
The equations are solved in the Carrington frame, in which we take the angular velocity of differential rotation to be
\begin{equation}
\omega(\theta) = 0.521 - 2.396\cos^2\theta - 1.787\cos^4\theta\,\,{\rm deg}\,{\rm day}^{-1}.
\end{equation}
The meridional circulation uses the \cite{2006A&A...459..945S} form
\begin{equation}
u_\theta(\theta) = u_0\sin(2\lambda)\exp\big(\pi - 2|\lambda | \big),
\end{equation}
where $\lambda=\pi/2 - \theta$ is latitude. In this article, we choose $u_0$ constant in time to give a peak flow speed of $11\,{\rm m}\,{\rm s}^{-1}$. We also choose a constant supergranular diffusivity of $\eta=500\,{\rm km}^2\,{\rm s}^{-1}$. These values were chosen to optimize agreement between the simulated and observed butterfly diagrams \citep[\textit{cf.}][]{2014SoPh..289..631Y}. For simplicity, we do not include any additional flux-decay term. The terms $S_\theta(\theta,\phi,t)$ and $S_\phi(\theta,\phi,t)$ in Equations (\ref{eqn:ath}) and (\ref{eqn:aph}) represent the emergence of new magnetic flux from beneath the photosphere. These terms are treated in a novel way, as described in Section \ref{sec:assim} below.

For this article, we simulate the period between Carrington Rotations  2074 and 2154 (namely 30th August 2008 and 17th September 2014). The observational input data used to generate both the initial $B_r$ map and the source terms $S_\theta(\theta,\phi,t)$, $S_\phi(\theta,\phi,t)$ are radial-component synoptic maps from GONG. These were chosen because they provide continuous coverage with consistent calibration over our period of interest, at a suitable resolution (360 pixels in $\phi$ and 180 pixels in $\sin\lambda$). For simplicity, we use a single synoptic map per Carrington rotation, with the most recent update at the end of the rotation. This time-resolution of input data is sufficient for present purposes. As already mentioned above, Figure \ref{fig:bfly} shows the resulting ``butterfly diagram'' of longitude-averaged $B_r$ on the photosphere, for both the observed magnetograms and the simulation.

\subsection{Assimilation Technique} \label{sec:assim}

Our treatment of the emerging flux terms $S_\theta(\theta,\phi,t)$ and
 $S_\phi(\theta,\phi,t)$ in Equations (\ref{eqn:ath}) and (\ref{eqn:aph}) differs from previous models. Typically, new flux is added either in the form of discrete bipolar magnetic regions \citep[\textit{e.g.}][]{1985SoPh...98..219S,2002SoPh..209..287M,2002ApJ...577.1006S,2006A&A...446..307B,
2007SoPh..245...87Y,2010ApJ...709..301J} or by assimilating observed magnetograms on the near-side of the Sun \citep{2000SoPh..195..247W,2003ApJ...590..493S,2004SoPh..222..345D,2010AIPC.1216..343A, 2014ApJ...780....5U}. The latter leads to more realistic reproduction of the observed photospheric field, but makes it difficult to isolate the contribution of individual emerging regions, as is our goal in this article. On the other hand, previous models using bipolar magnetic regions have assumed an idealized form for the newly emerging regions that is limited in accuracy for complex active regions. As a result, we employ here a new technique for assimilating individual strong-flux regions into the simulation. The aim is to separate individual new regions, while maintaining their detailed structure from the observed magnetogram.

The assimilation algorithm is described in detail in Appendix \ref{sec:algorithm}. To illustrate, Figure \ref{fig:regions} shows the strong-flux regions identified by our numerical code for CR 2132. As can be seen, the algorithm identifies both isolated bipolar magnetic regions and multipolar complexes of activity. The regions are incorporated one-by-one into the simulation on their day of central-meridian crossing.

\begin{figure}[h]
\centering
\includegraphics[width=0.8\textwidth]{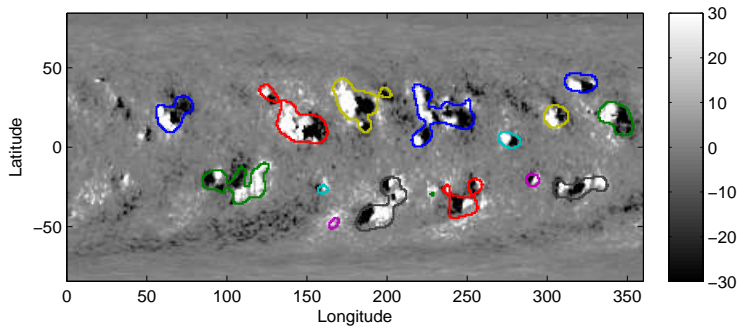}
\caption{Strong-flux regions identified by the automated technique for the GONG synoptic map of CR 2132. Region outlines are shown in different colours, superimposed on the observed synoptic map of $B_r(\theta,\phi)$ (saturated at $\pm 30\,{\rm G}$).}
\label{fig:regions}
\end{figure}

There are two major advantages of this procedure over previous methods for assimilation of purely bipolar regions. Firstly, it is fully automated, giving consistency and objectivity in selection of new flux regions, as well as saving a considerable amount of time. Secondly, the new strong-flux regions \emph{replace} the pre-existing simulated flux within those regions, rather than being superimposed. This is a major advantage because it allows the model to incorporate subsequent flux emergence observed over multiple solar rotations within existing activity complexes. In the bipole-based emergence model \citep[\textit{e.g.}][]{2014SoPh..289..631Y}, this is very difficult to model realistically. Yet the structure of these activity groups plays a major role in the global distribution of magnetic flux on the Sun.

\subsection{Grouping of New Regions}

For the analysis in this article, we make the additional step of grouping together new regions when they emerge at the same location in space, separated in time by at most two Carrington rotations. This is necessary because the assimilation algorithm often modifies an already emerged region, if it still has strong flux in a subsequent rotation. Each group should therefore be considered together when assessing its contribution to a polar surge or the polar field. Of the 204 strong-flux regions identified between CR 2074 and CR 2116, we identify 96 separate groups. The groups range in size from 1 to 15 regions, with a mean size of 2.125 regions, but large groups are rare and the median size is only 1 region.

\section{Origin of Poleward Surge A}  \label{sec:source}

To determine which of the emerged-flux groups produced polar surge A, we repeated the flux-transport simulation omitting each of the groups in turn. We then quantify the effect of each group by cross-correlating the butterfly diagram from each run against that from the complete run, in a window from CR 2104 to CR 2117 and $\lambda=35^\circ$ to $\lambda=70^\circ$. This window is chosen to pick out surge A.

\begin{figure}[h]
\centering
\includegraphics[width=0.8\textwidth]{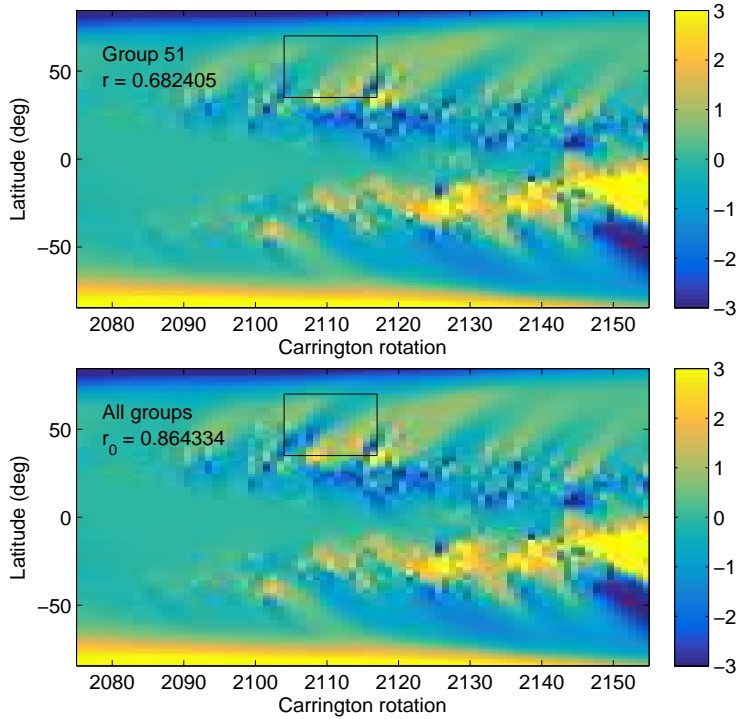}
\caption{Effect of omitting group 51 from the flux-transport simulation, shown in longitude-averaged $B_r$. The top panel shows the run with group 51 omitted, while the bottom panel shows the complete run. The cross-correlations $r$ and $r_0$ are computed in the indicated window, against the observed GONG butterfly diagram.}
\label{fig:bfly51}
\end{figure}

For example, Figure \ref{fig:bfly51} shows the butterfly diagram when group 51 is omitted, compared to that from the complete run. The window cross-correlation between the complete run and the observed butterfly diagram (Figure \ref{fig:bfly}a) is $r_0=0.864$. This residual error must result from processes that are not fully accounted for in our simple flux-transport model. By contrast, the run with group 51 omitted has a lower cross-correlation, $r=0.682$, indicating that group 51 has a significant effect on surge A.

\begin{figure}[h]
\centering
\includegraphics[width=\textwidth]{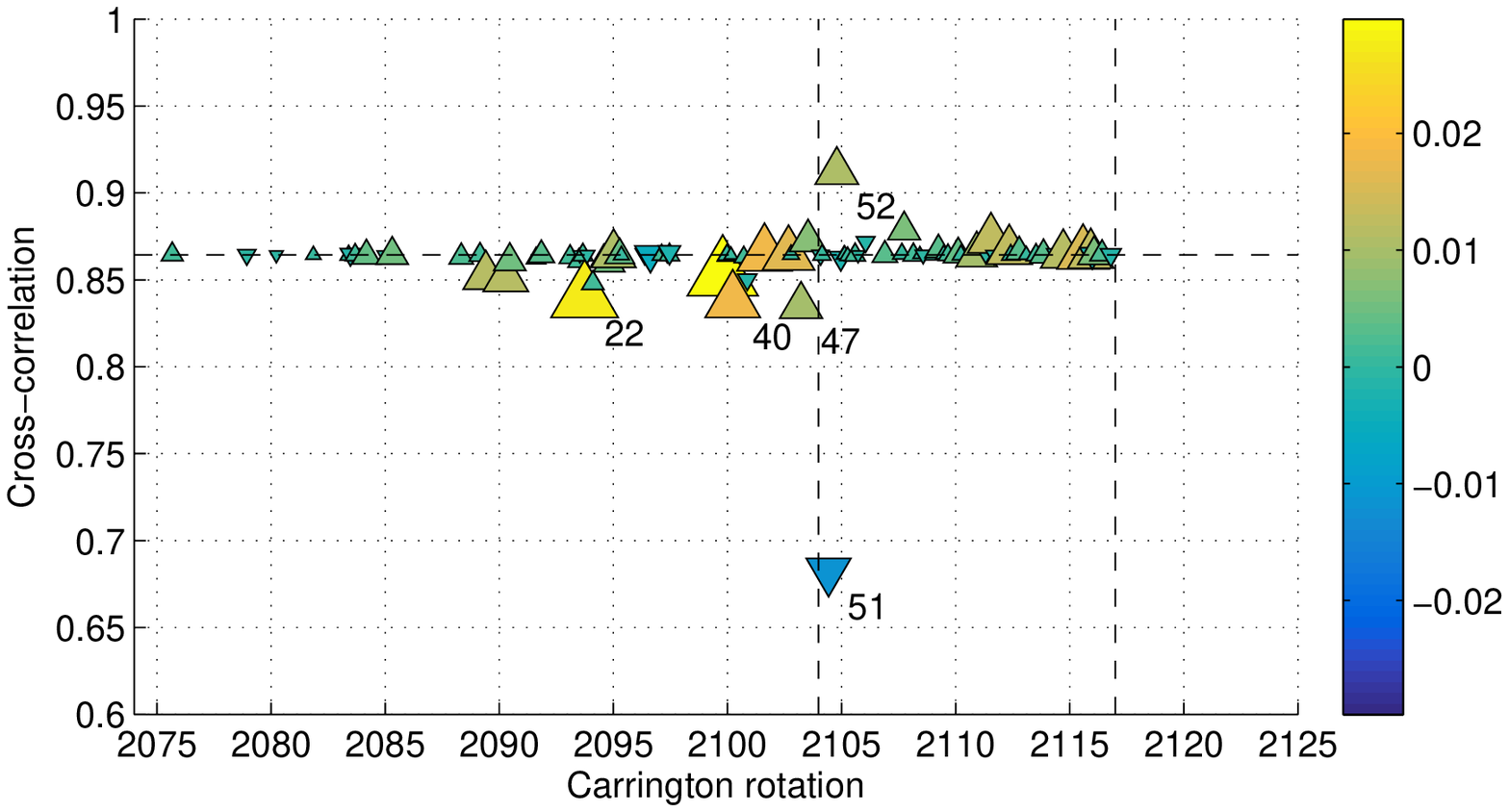}
\caption{Cross-correlation between the observed and simulated butterfly diagrams. Each triangle corresponds to a simulation with a different group of regions omitted, located at the time of insertion of the first region in that group. The colour and size of the triangle are proportional to the (average) axial dipole moment [${\rm G}$] among that group of inserted regions, as indicated in the colourbar, with upward triangles for positive values and downward for negative. Dashed-vertical lines show the extent of the correlation window, while the dashed-horizontal line shows the cross-correlation [$r_0$] when all groups are included.} \label{fig:bfly_cor}
\end{figure}

In fact, the correlation analysis shows that group 51 is primarily responsible for surge A. To see this, Figure \ref{fig:bfly_cor} shows the cross-correlation coefficient [$r$] for all groups (up to and including CR 2116). The colour and size of each symbol denote the average axial dipole moment of each group, to be discussed below. Focusing for now on the $r$-values, it is seen that omitting the majority of groups has no effect on the correlation in this window. Only groups 22, 40, 47, 51, and 52 cause $r$ to differ from $r_0$ by more than $0.02$, and only group 51 causes a difference of more than $0.05$. For reference, the NOAA active-region numbers associated with these five groups are given in Table \ref{tab:noaa}.

\begin{table}
\caption{NOAA active-region numbers for the groups affecting the correlation window.
}
\label{tab:noaa}
\begin{tabular}{ccc}     
  \hline                   
  Group & Carrington rotations & NOAA regions\\
  \hline
22 & CR 2093 & 11048\\
40 & CR 2100 & 11097\\
47 & CR 2103 & 11120\\
51\tabnote{Illustrated in Figure 5.} & CR 2104\,--\,5, CR 2107\,--\,8 & 11131, 11140, 11164, 11180\\
52 & CR 2104, CR 2106\,--\,11 & 11135, 11157, 11169, 11185\,--\,6, 11189,\\
& & 11203\,--\,5, 11228\,--\,30, 11242\\
  \hline
\end{tabular}
\end{table}

Before considering group 51 in more detail, let us remark on the other five groups affecting our correlation window. Groups 22, 40, and 47 each consist of only a single bipolar region, tilted in accordance with Joy's law. These regions contribute to the cross-correlation primarily by strengthening the positive flux region in the top left of the correlation window, although their leading polarity flux does contribute to the negative surge A. Group 52 is notable in that omitting it actually increases $r$ (\textit{i.e.} improves the match with the observed butterfly diagram). This group is a large cluster of multiple bipolar regions, persisting over seven rotations (CR 2104 to CR 2111). It is responsible for much of the positive surge in the bottom-right of the correlation window. Compared to the observations, this surge is too strong; thus omitting group 52 actually improves the correlation. This suggests that the processes in our flux-transport model may be overestimating the poleward spreading from large complexes such as group 52, perhaps because we neglect the effect of  systematic inflows towards active regions \citep{2006ESASP.624E..12D,2010ApJ...717..488B,2010ApJ...717..597J}.

\begin{figure}[h]
\centering
\includegraphics[width=\textwidth]{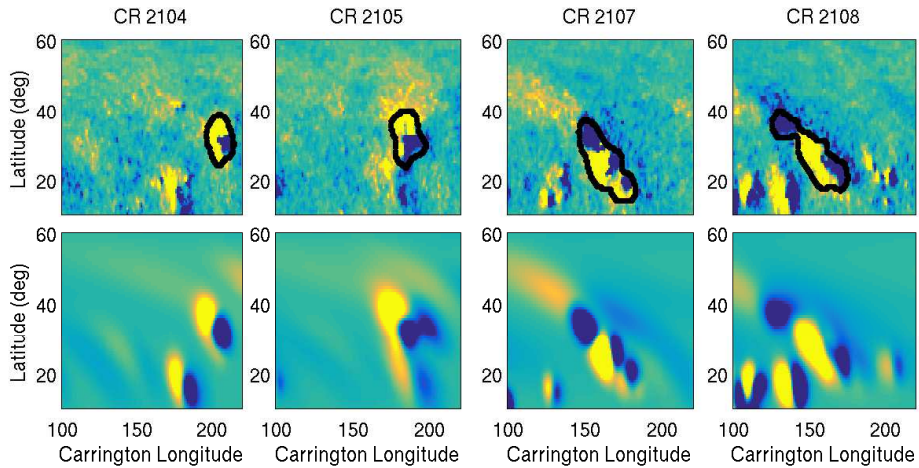}
\caption{The four strong-flux regions comprising group 51, shown outlined in the GONG synoptic magnetograms (top row) and simulated $B_r$ (bottom row). The simulated $B_r$ is shown at the end of the corresponding Carrington Rotation. All images are saturated at $\pm 15\,{\rm G}$, yellow positive and blue negative.} \label{fig:group51}
\end{figure}

The majority of the negative flux in the poleward surge clearly arises from group 51. This group consists of four individual strong-flux regions, determined from the synoptic maps for CR 2104, CR 2105, CR 2107, and CR 2108. These are highlighted in the synoptic maps in Figure \ref{fig:group51}, above the simulated field (at the end of the corresponding Carrington rotation). It is seen that group 51 consists of i) a bipolar active region appearing first in CR 2104, and still having strong flux in CR 2105, and ii) a multipolar region appearing first in CR 2107 and still visible in CR 2108. Both of these constituent regions appear to have a large tilt angle, which is understood as the angle between their positive and negative centroids with respect to the East-West line, although they are oppositely tilted. Since the later region is multipolar in nature, a better way to quantify this is the axial dipole moment of the region, defined as
\begin{equation}
b_{1,0}=\frac{3}{4\pi}\int_{R_i}B_r(\theta,\phi)\cos\theta\,\mathrm{d}\Omega.
\end{equation}
The four strong-flux regions comprising group 51 have $b_{1,0}$ of $0.0120\,{\rm G}$, $0.0182\,{\rm G}$, $-0.0546\,{\rm G}$, and $-0.0205\,{\rm G}$ respectively. Hence the group average is negative ($-0.0112\,{\rm G}$), as shown in Figure \ref{fig:bfly_cor}. We argue that this relatively large and predominantly negative dipole moment of the later region in group 51 (opposite to the majority positive sign from Joy's law) is primarily responsible for surge A in the butterfly diagram. The effect is strengthened by the relatively high latitude at which this region emerges. However, there is also a smaller contribution from the leading polarity of the earlier bipolar region in the group, and to a lesser extent from groups 22, 40, and 47.

\section{Contribution to the Polar Field} \label{sec:polar}

It would be erroneous to suppose that, merely because group 51 causes a particularly prominent opposite-polarity polar surge in the butterfly diagram, it must make a particularly strong contribution to the net polar field. In fact, since group 51 is located at rather high latitude, the flux transport model predicts that its net polar field contribution at the end of Cycle 24 will be weak. The strong poleward surge A seen in the butterfly diagram is merely a transient phenomenon. This is well known from previous studies of surface-flux transport \citep{1991ApJ...375..761W, 2002SoPh..207..291M}, but it is interesting to illustrate this for our specific region 51.

\begin{figure}[h]
\centering
\includegraphics[width=0.8\textwidth]{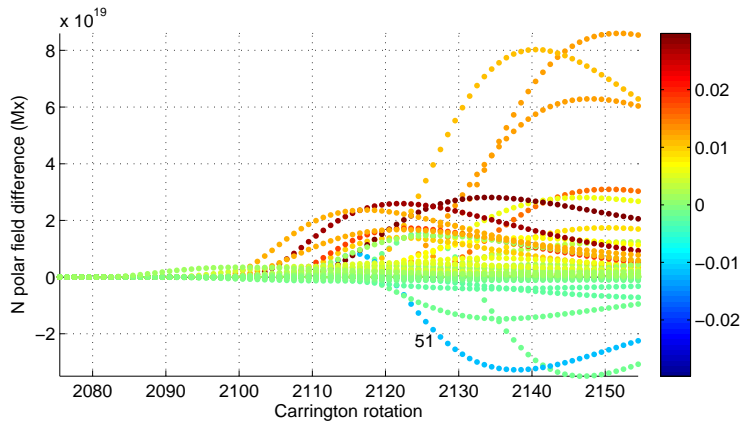}
\caption{Change in north-polar flux [$\Phi_{\rm NP}(t)$] compared to the full simulation, in the runs where each emerging group is omitted. The curves are coloured according to the average axial dipole moment [$b_{1,0}$] of the group [${\rm G}$]. The curve for group 51 is labelled.} \label{fig:polar}
\end{figure}

To see the relative effect of the different emerging groups on the polar field, we show in Figure \ref{fig:polar} the change in the north polar flux 
\begin{equation}
\Phi_{\rm NP}(t) = R_\odot^2\int_{\theta < 20^\circ}B_r(\theta,\phi,t)\,\mathrm{d}\Omega
\end{equation}
as each group is omitted from the simulation. It is seen that the largest changes of polar flux come from a relatively small number of emerging groups, and many of these are the groups with largest $|b_{1,0}|$. (The others are groups with many constituent regions, but weak average dipole moment.) Additionally, the sign of the difference in Figure \ref{fig:polar} is seen to correlate with the sign of $b_{1,0}$ for each group. The majority of groups have $b_{1,0}>0$ (in accordance with Joy's law for Cycle 24), and therefore contribute positively to $\Phi_{\rm NP}$. Group 51 is conspicuous in its negative $b_{1,0}$, and one of few groups to make a net negative contribution to $\Phi_{\rm NP}$.

A crucial observation in Figure \ref{fig:polar} is that the polar-field contribution from many groups -- including group 51 -- appears to initially peak, then to decay in time. This is entirely typical: the initial peak is caused by the negative surge reaching the pole first, and then the decay is caused by the positive surge from the same region reaching the pole at a later time.

\begin{figure}[h]
\centering
\includegraphics[width=0.8\textwidth]{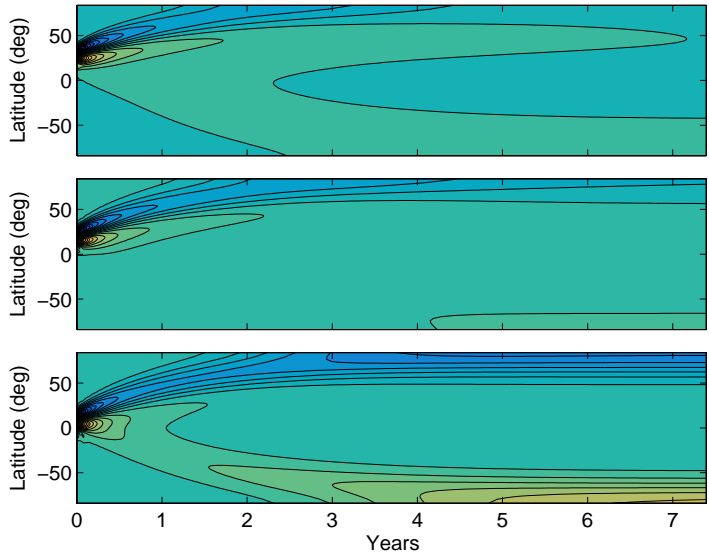}
\caption{Longitude-averaged $B_r$ in the three simulations of an identical region emerged at different latitudes. (Yellow shows positive, blue negative).} \label{fig:testbfly}
\end{figure}

\begin{figure}[h]
\centering
\includegraphics[width=0.8\textwidth]{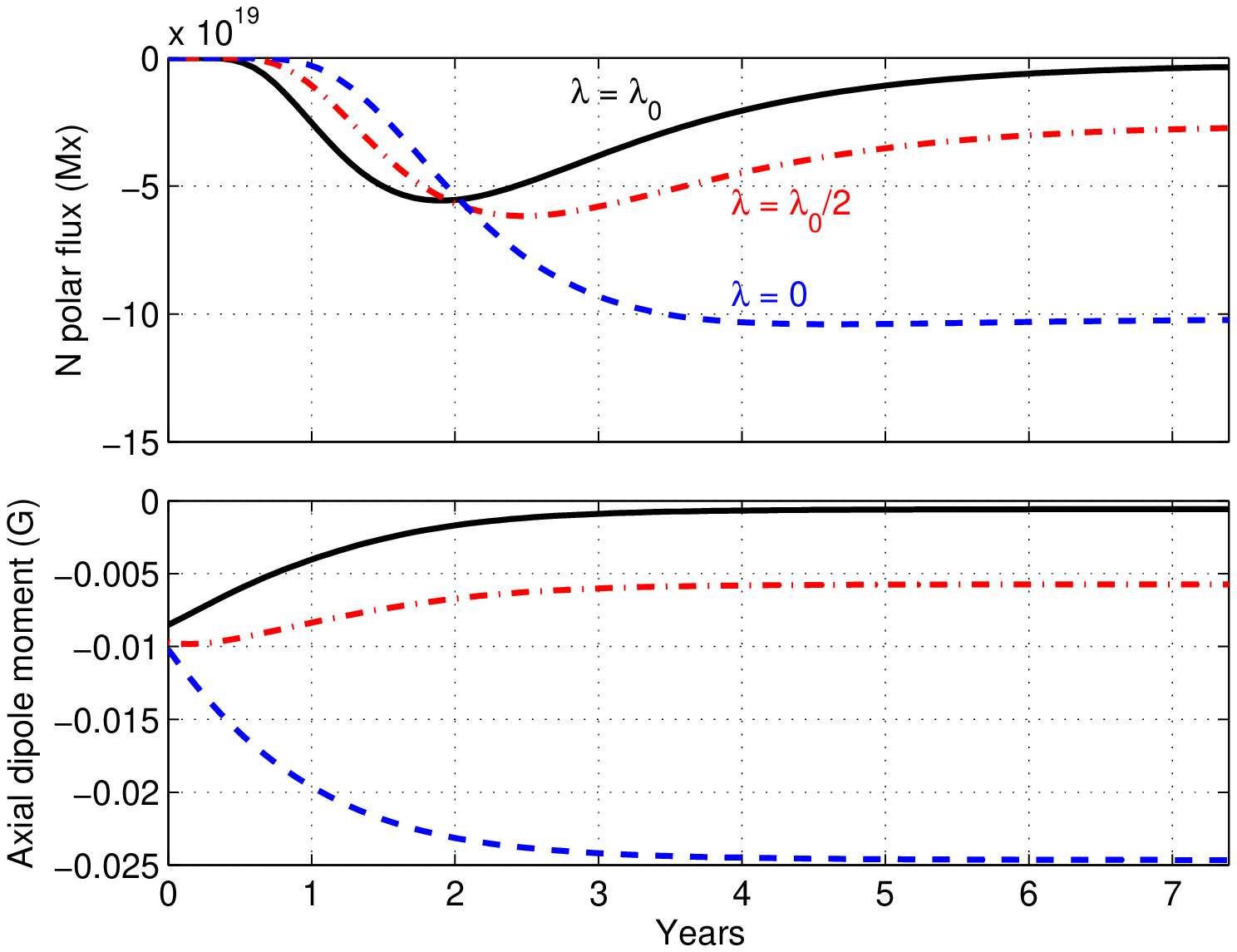}
\caption{North polar flux [$\Phi_{NP}(t)$] and axial dipole moment [$b_{1,0}(t)$] as a function of time for the three simulations of an identical region emerged at different latitudes.} \label{fig:testpolar}
\end{figure}

It is cleanest to illustrate this effect by simulating the evolution of a single region in isolation, without the surrounding field \citep[\textit{e.g.}][]{2002SoPh..207..291M}. We choose the strong-flux region emerging in CR 2107 (\textit{i.e.} in the third column of Figure \ref{fig:group51}). Figure \ref{fig:testbfly} shows three butterfly diagrams obtained in such single region simulations, when the complex is placed i) at its observed latitude $\lambda=\lambda_0\equiv 24^\circ{\rm N}$, ii) at $\lambda=\lambda_0/2$, and iii) at $\lambda=0$. For the same three runs, Figure \ref{fig:testpolar} shows both the polar flux [$\Phi_{\rm NP}(t)$] and the axial dipole [$b_{1,0}(t)$] as functions of time. The latter can be compared to Figure 6 of \cite{2014ApJ...791....5J} and Figure 4 of \cite{1991ApJ...375..761W}.

From Figures \ref{fig:testbfly} and \ref{fig:testpolar} we see the following behaviour:
\begin{enumerate}[i)]
\item The polar flux does not (in general) decay to zero, but reaches an asymptotic steady state after about six years, which is symmetric between the two hemispheres. \citep[This results from a balance between equatorward diffusive transport and poleward meridional flow;][]{1984SoPh...92....1D}.
\item Placing the emerging region nearer the Equator leads to a larger asymptotic value of the polar field. (The exact latitudinal dependence depends on the chosen meridional-flow profile and supergranular diffusivity).
\item For the latitudes shown, the axial dipole moment does not show the turnover seen in the polar field for $\lambda_0$ and $\lambda_0/2$. This is because it does not have to wait for flux to be transported to the pole to register the presence of the region. For the same reason, it tends to converge more rapidly to its asymptotic value than does the polar flux, although the asymptotic values of the two quantities are proportional.
\end{enumerate}

It is clear that, despite its significant contribution to the transient poleward surge A, group 51 is likely to make a relatively small contribution to the polar field at the end of Cycle 24 owing to its high latitude of emergence. The key point is that a decaying active region will produce a net (asymptotic) polar field only if it leads to a net imbalance in magnetic flux in each hemisphere \citep{1985AuJPh..38.1045G,1986PhDT.........3D,2002ApJ...577L..53W}. This can only happen if net flux from the region either emerges on both sides of the Equator, or is transported across it. It follows that those active regions located near the Equator -- and in particular straddling the Equator -- are the most important for producing fluctuations in the polar-field production process \citep[\textit{cf.}][]{2013A&A...557A.141C}.

\section{Conclusions}

In summary, the activity group responsible for poleward surge A is a striking example of the statistically fluctuating nature of active-region flux emergence on the Sun. This group has a large axial dipole moment, opposite in sign to Joy's law. It is the primary source of the poleward surge A that is clearly seen in the magnetic butterfly diagram.

From the viewpoint of the Babcock--Leighton model for the solar cycle, the importance of an individual emerging region lies in its net contribution to the polar field. Our modelling shows that this particular activity group, whilst causing a significant transient perturbation to the polar field, will not lead to a large long-term contribution. This is because it is located far from the Equator, and it is the cross-equatorial flux transport that controls the asymptotic polar field \citep[as was first recognised by][]{1985AuJPh..38.1045G}. On the other hand, the transient perturbation does last for about six years. This suggests that higher-latitude regions emerging later in the cycle may still be significant contributors to polar field measurements taken at a fixed time late in the cycle. It is also not clear that the best surface indicator of underlying poloidal magnetic field in the convection zone will be the polar field alone \citep{1991ApJ...375..761W}. The sub-surface decay of active-region flux at lower latitudes may well be an important factor in repairing the azimuthal field belts \citep{2007ApJ...659.1713V, 2013MNRAS.436.3366Y}. In this case, the role of high-latitude active regions in the Babcock--Leighton process may be more significant than their polar-field contribution might suggest.

Finally, we have developed a new technique for data-assimilation in the flux transport model. This demonstrates the feasibility of a middle ground between inserting idealised bipolar regions and assimilating full magnetograms. In the future this could be combined with more sophisticated automated techniques for detecting and inserting active regions at higher time cadence \citep[\textit{e.g.}][]{2011AdSpR..47.2105H}.

\begin{acks}
We acknowledge the Leverhulme Trust for funding the `'Probing the Sun: inside and out'' project upon which this research is based. ARY thanks STFC for financial support through consortium grant ST/K001043/1. The research leading to these results has received funding from the European Union's Seventh Programme for Research, Technological Development and Demonstration under Grant Agreement No. 284461 (eHEROES project). LvDG acknowledges the Hungarian government for grant OTKA K 109276. DB and LvDG thanks STFC for support under Consolidated Grant ST/H00260/1. This work utilizes data obtained by the Global Oscillation Network Group (GONG) program, managed by the National Solar Observatory, which is operated by AURA, Inc. under a cooperative agreement with the National Science Foundation. The data were acquired by instruments operated by the Big Bear Solar Observatory, High Altitude Observatory, Learmonth Solar Observatory, Udaipur Solar Observatory, Instituto de Astrof\'{\i}sica de Canarias, and Cerro Tololo Interamerican
Observatory. We thank Duncan Mackay for reading an initial draft, and the referee for helpful suggestions.
\end{acks}


\appendix   

\section{Assimilation Algorithm} \label{sec:algorithm}

To assimilate new magnetogram flux, we first process each individual synoptic magnetogram $g(\theta,\phi)$ to determine a list of strong-flux regions, using the following procedure:
\begin{enumerate}[i)]
\item Correct any overall net flux imbalance in $g$ by subtracting an equal amount from each pixel.
\item Take the absolute value to obtain a new map $f(\theta,\phi)=|g(\theta,\phi)|$.
\item Smooth $f$ with a Gaussian filter (standard deviation $\sigma$) to obtain $\bar{f}$. The purpose of this step is primarily to merge together the positive and negative polarities of individual bipolar active regions.
\item Identify connected regions [$R_i$] where $\bar{f}>B_0$. These are the new regions that will be inserted into the flux-transport model. (Note that these regions need not be simply connected and could have holes.)
\end{enumerate}

Next, each strong flux region [$R_i$] is incorporated into the flux-transport model as follows:
\begin{enumerate}[i)]
\item Determine the centroid of $R_i$ in Carrington longitude, and choose to insert it on the corresponding day when this longitude crosses central meridian.
\item Determine the net flux imbalance $\Phi_i=\int_{R_i}B_r\,\mathrm{d}\Omega$, where $B_r(\theta,\phi)$ is the pre-existing \emph{simulated} radial field.
\item Modify the observed field [$g|_{R_i}$] so that the net flux in $R_i$ matches $\Phi_i$, by subtracting an equal amount from each pixel. This prevents us introducing a flux imbalance in the insertion step.
\item Replace the simulated $B_r$ in $R_i$ with the observed field [$g|_{R_i}$]. 
\item Recompute the global vector potential [$A_\theta,A_\phi$] to match the modified $B_r$.
\end{enumerate}

The process is designed to be fully automated, once the controlling parameters $\sigma$ and $B_0$ have been determined. For the GONG synoptic maps, we have found $\sigma=3$ pixels and $B_0=15\,{\rm G}$ to work well. The smoothing width $\sigma$  should be large enough to merge the two opposite polarities of activity complexes, while not merging too many neighbouring complexes. The threshold $B_0$ needs to be large enough to include all significant new flux, but be above the ``noise'' level of ephemeral regions (in the smoothed magnetogram), so that only flux at active latitudes is assimilated.

One must be aware that, in order to preserve the field distribution, the net flux in each identified region $R_i$ should be close to that existing beforehand in the simulation. For example, a region emerging in an empty area of the solar surface should have zero net flux. Otherwise, the flux correction will noticeably distort the observed field. This practical consideration requires $\sigma$ to be sufficiently large and $B_0$ to be sufficiently small.




\end{article} 

\end{document}